\documentclass[aps,twocolumn,superscriptaddress,eqsecnum,nofootinbib,pre]{revtex4-1}

\usepackage{scrextend}
\usepackage{amssymb}  
\usepackage{amsmath}   
\usepackage{natbib}
\usepackage{epsfig}   
\usepackage{relsize}
\usepackage{graphicx,subfigure}   
\usepackage{dcolumn}
\usepackage{bm}  
\usepackage[english]{babel}
\usepackage[latin1]{inputenc}
\usepackage{hyperref}
\usepackage{xcolor}
\usepackage{comment}
\usepackage{nameref}
\usepackage[update]{epstopdf}
\emergencystretch=\maxdimen
\begin{document}

% Title must be 250 characters or less.

\title{Connecting metapopulation heterogeneity to aggregated lifetime statistics}

\author{E. H.  Colombo} 
\email{ecolombo@ifisc.uib-csic.es}
\affiliation{IFISC (CSIC-UIB), Campus Universitat Illes Balears, 07122, Palma de Mallorca, Spain}

% Please keep the abstract below 300 words
\begin{abstract}
Aggregated metapopulation lifetime statistics has been used to access stylized facts that might help identify the underlying patch-level dynamics. For instance, the emergence of scaling laws in the aggregated probability distribution of patch lifetimes can be associated to critical phenomena, in which the correlation length among system units tends to diverge. Nevertheless, an aggregated approach is biased by patch-level variability, a fact that can blur the interpretation of the data. Here, I propose a weakly-coupled metapopulation model to show how patch variability can solely trigger qualitatively different lifetime probability distribution at the aggregated level. In a generalized approach, I obtain a two-way connection between the variability of a certain patch property (e.g. carrying capacity, environment condition or connectivity) and the aggregated lifetime probability distribution. Furthermore, for a particular case, assuming that scaling laws are observed at the aggregated-level, I speculate the heterogeneity that could be behind it, relating the qualitative features the variability (mean, variance and concentration) to the scaling exponents. In this perspective, the application points to the possibility of equivalence between heterogeneous weakly-coupled metapopulations and homogeneous ones that exhibit critical behavior.
\end{abstract}

\maketitle

\section{Introduction}

A metapopulation is composed of subpopulations that live within disconnected habitats (patches) but can interact through individual dispersal. Locally, at each patch, due to demographic stochasticity, populations are vulnerable to extinction. However, the metapopulation as a whole tends to be stable as individual dispersal can balance fluctuation and recolonize patches~\cite{HanskiBook}. This interplay between local and nonlocal dynamics generates, for each patch, a collection of time intervals that correspond either to an empty or occupied state. With this dataset statistical analysis can be performed to extract relevant information about the metapopulation dynamics~\cite{natureHanski2000,HanskiBook,tilman1994,kuussaari2009}.

Aggregated approaches (i.e., grouping data from all patches and species) has been used to access stylized facts, providing a deepen view of inter-patch interaction. Previously, the existence of power-law scaling in the aggregated distribution of species abundance and lifetimes (the occupied time interval) has been associated with critical phenomena that generates interaction between system units across a broad range of scales~\cite{mixing,stanleybirds,stanleybook,sole1999}. Nevertheless, it has been also pointed out that heterogeneity could mimic similar statistical features, a fact that can compromise the interpretation of the underlying dynamics~\cite{mixing}.

\begin{figure}
\includegraphics[width=0.8\columnwidth]{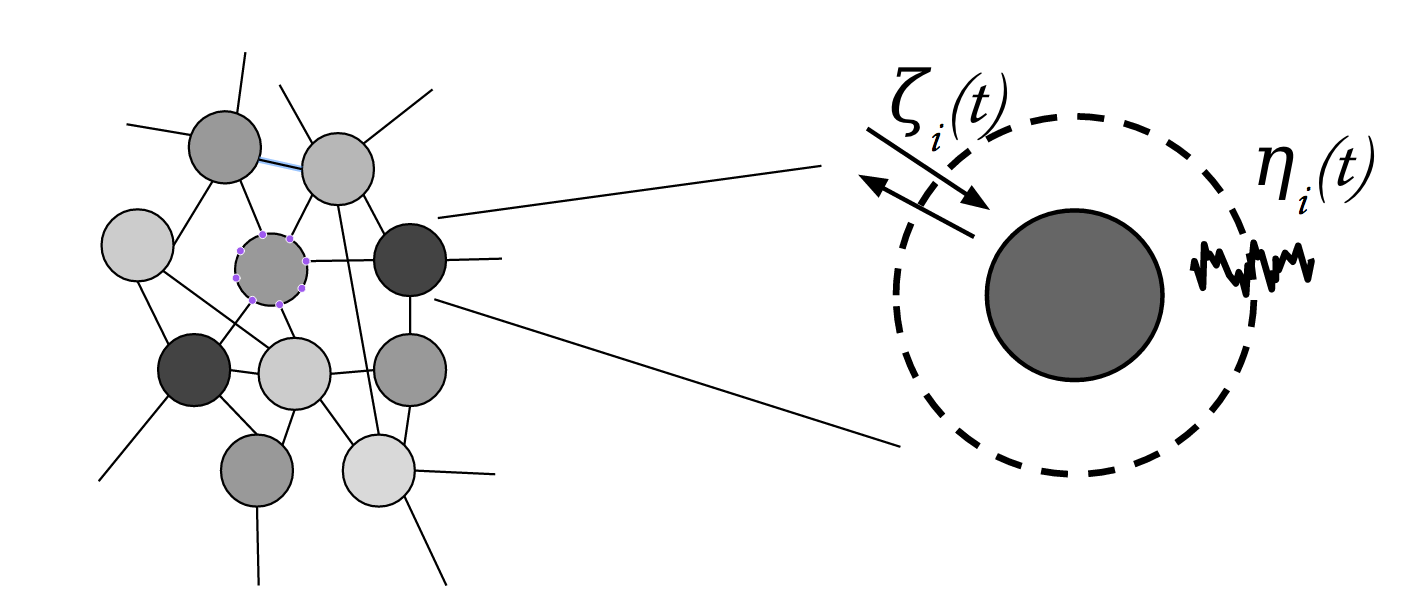}
\caption{The weakly-coupled metapopulation model. The dynamics of each patch $i$ is considered to be independent and includes environmental noise $\eta_i(t)$ and migration flux $\zeta_i(t)$ terms to effectively take into account the coupling among patches and environment (see supplementary material for examples).}
\label{fig:scheme}
\end{figure}

This work deepens previous discussions on the role of heterogeneity in aggregated datasets promoted in Ref.~\cite{mixing}. Focusing on the aggregated lifetime statistics, the following results provide an explicit analytical relation between patch-level variability and the aggregated lifetime probability distribution. For that, I propose a weakly-coupled metapopulation model, in which dispersal timescales are longer than population dynamics correlation times. This way, each patch dynamics is constructed by a local part and an effective term to account for the flux of individuals from other patches (see Fig.~\ref{fig:scheme}). In this framework, in Sec.~\ref{sec:dataset}, I start by characterizing the structure of the lifetime dataset. In Sec.~\ref{sec:superstat}, the aggregated lifetime probability distribution is written as a superstatistics of the local dynamics~\cite{superstatistics}, i.e. by mixing each patch exponential distribution with different characteristic timescales~\cite{mixing}.
Finally, in Sec.~\ref{sec:deconvo}, I obtain the reverse transformation that can reveal patch heterogeneity from the aggregated lifetime probability distribution. Overall, even though, locally, lifetimes are exponentially distributed, due to the variability in the patches' properties (such as its carrying capacity, environmental condition and connectivity), the aggregated lifetime probability distribution could exhibit forms that deviate from the exponential form~\cite{superstatistics,mixing}.

In Sec.~\ref{sec:apply}, by applying the model, I speculate the patch variability behind a generalized power-law form for the aggregated lifetime distribution. This particular form is inspired by the North American Breeding Bird Survey dataset~\cite{databirds,stanleybirds} for which was invoked the presence of critical phenomena to justify the scaling laws~\cite{mixing}. Thus, these applied results aim to show the possible equivalence between heterogeneous weakly-coupled metapopulations and homogeneous ones that exhibit critical behavior at the aggregated-level. The relation between the scaling exponents with the qualitative features of the patch-level variability (mean, variance and concentration) is explored in detail.

The idea of connecting patch and global levels has been a central issue in metapopulation dynamics studies \cite{HanskiBook,natureHanski}. However, a rigorous mathematical treatment for stochastic spatial explicit models faces challenges in providing expressions for macroscopic quantities. Despite recent developments on approximative methods, such as moment closure or perturbative expansions around the mean-field solution, only superficial aspects of the system are straightforward to access~\cite{OvaskainenFrame,otso1}. The presented weakly-coupled model bypass these challenges, allowing to demonstrate the emergence of scaling purely due to heterogeneity. This is a necessary reference frame to understand the ecosystem across scales~\cite{localglobal,lafuerza2013} and to ensure a proper identification of emergent behavior caused by ecosystem variability (its landscape structure and species), which can play a role in its resilience, stability and functions~\cite{Loreau2001,Loreau2003,sole1999,stanleybook}.

\section{Framework}
\label{sec:framework}
In the following sections, I start by characterizing the lifetime dataset structure and establishing the direct transformation from patch to the ensemble (aggregated) level, obtaining the lifetime probability distribution considering data from all patches. Then, I find the inverse transformation that reveals patch variability from the aggregated perspective. Later in Sec.~\ref{sec:apply}, I apply the approach assuming specific population dynamics models and discuss the appearance of power-law statistics.

\subsection{Dataset structure}
\label{sec:dataset}

At patch-level, stochastic fluctuations (demographic and environmental) eventually drive the population to the null (extinction) state, which is locally-absorbing~\cite{extinction}. However, since each patch's boundary is open, new individuals are able to reestablish the population. The combination of these two forces places the patches out of equilibrium in a succession of colonization and extinction events. 

Experimental records for a metapopulation constituted by $N$ patches can be structured as
\begin{equation}
\label{patchset}
\mathcal{E}_i = \{T^{\dagger}_{i,0},T_{i,0},T^{\dagger}_{i,1},T_{i,1},\ldots\}\, ,
\end{equation}
where $i\in (1,2,\ldots,N)$, $T_i$ is the time interval in which the patch $i$ was occupied, its lifetime, and $T^\dagger_i$ represent the time interval in which the patch was empty~\cite{hanski}. 

For each patch, the number of records $n_i \equiv | \mathcal{E}_i |$ is given by
\begin{equation}
n_i \sim \frac{T_o}{ \langle T_i \rangle + \langle T^{\dagger}_i \rangle }\, .
\label{number}
\end{equation}
where $T_o$ is the total observation time.

\begin{figure}[h]
\includegraphics[width=\columnwidth]{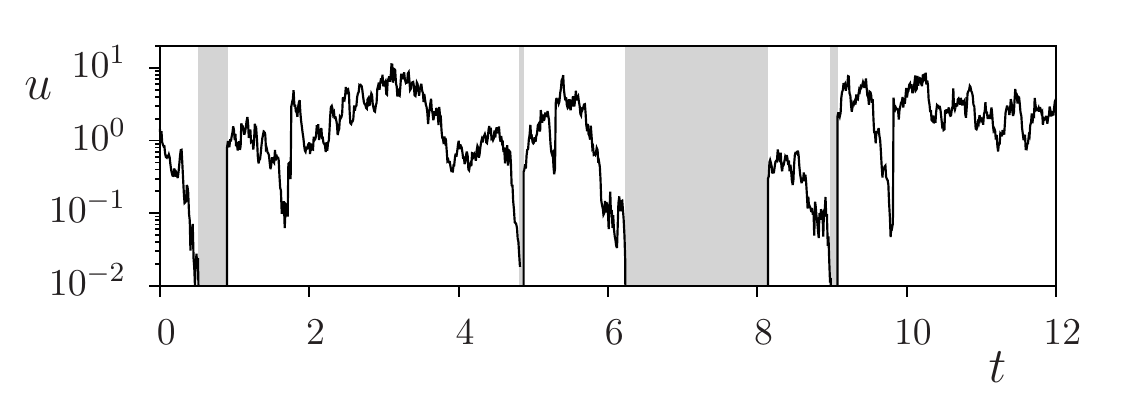}
\caption{Typical time-series for the population density $u$ at a certain patch. Solid lines represent results from the canonical effective model as described in supplementary material. Gray regions depict the time windows in which the patch was empty, $u=0$. For the values of the parameter chosen the patch mean lifetime $\tau\simeq 13.23$ and $\tau^\dagger=1.0$.}
\label{fig:series}
\end{figure}

Since patches with same features will exhibit same statistical trends,
we can rewrite Eq.~(\ref{number}), defining the mean lifetime as a function of the property $m$,
\begin{equation}
n(m) \sim \frac{N(m)T_o}{\tau(m) +  \tau^\dagger(m)}\, .
\label{number2}
\end{equation}
$n(m)$ gives the total number of events for all patches with property $m$ within the observation time, being $N(m)$ the number of patches with property $m_i \in [m_i-\Delta m, m_i+ \Delta m]$ with $\Delta m/m \ll 1$. In Fig.~\ref{fig:series}, for a particular model that we will discuss later (see also supplementary material) a typical time series is depicted, highlighting patches' occupied and empty time intervals.

From the aggregated perspective, we can see that patches with small lifetimes have a higher importance in the statistics, due to large event counting. Besides that, these weights are modulated by the likelihood of having a particular patch property in the metapopulation. Thus, in the large metapopulation limit, $N \gg 1$, the weight $w(m)$ of a particular property $m$ is given by
\begin{equation}
\label{weight}
w(m) =  \frac{\mathcal{N}\rho(m)}{\tau(m) +  \tau^\dagger(m)}\, ,
\end{equation}
where $\rho(m) = \frac{N(m)}{N}$ is the probability density function for observing a certain property $m$ in the metapopulation and $\mathcal{N}^{-1} = \int_\Omega \frac{\rho(m)}{\tau(m) +  \tau^\dagger(m)} dm$ is a normalization factor.

\subsection{Superstatistics}
\label{sec:superstat}
In order to construct the aggregated lifetime probability distribution $\mathcal{P}(T)$, it is necessary to mix those of the patches' accounting for the contributions of different local properties. Namely, in a given domain $\Omega$, 
\begin{equation}
\label{superstat}
\mathcal{P}(T) = \int_\Omega w(m)p(T| m) dm\, .
\end{equation}
where $p(T|m)$ is the lifetime probability distribution for a patch having characteristic $m$ and $w(m)$ is the weight of the characteristic in the aggregated set.
Substituting the weight given by Eq.~(\ref{weight}) in (\ref{superstat}),
\begin{equation}
\mathcal{P}(T) = \mathcal{N}\int_\Omega \frac{\rho(m)}{\tau(m) +  \tau^{\dagger}(m)} p(T| m) dm\, .
\end{equation}

It has been shown that lifetimes are, rather generally, exponentially distributed independently of the model specificities~\cite{volker}. In the framework of Markov processes, being $q(n,t)$ the probability of having $n$ individuals at instant $t$, its temporal evolution is given by $\dot{q} = M(q)q$, where matrix $M$ sets the transition rates for increasing ($n\to n+1$) and decreasing ($n\to n-1$) the number of individuals. Typically, it is found that the leading eigenvalue $\Lambda_1$ of $M$ is much smaller than the others, even when including density dependency, environment noise and other features~\cite{volker}. As a consequence, after a fast transient $q_{n>0}(t) \propto e^{-\Lambda_1 t}$ the extinction probability becomes $q_0(t) \propto 1 - A e^{-\Lambda_1 t}$. Hence, the lifetime probability distribution is given by $p \propto e^{-\Lambda_1 t}$ with mean lifetime $\tau \simeq 1/\Lambda_1$~\cite{volker}. In the supplementary material, I investigate in detail the lifetime distribution for the canonical model~\cite{extinction}. There, exponential laws were also found.

Motivated by this feature, setting $p(\tau|m) = e^{-T/\tau(m)}/\tau(m)$ in Eq.~(\ref{superstat}), we obtain
\begin{equation}
\mathcal{P}(T) = \mathcal{N}\int_\Omega \frac{\rho(m)}{\tau(m)[\tau(m) +  \tau^{\dagger}(m)]} e^{-T/\tau(m)} dm\, .
\label{predirect}
\end{equation}

Eq.~(\ref{predirect}) defines the aggregated lifetime probability $\mathcal{P}$ as a mixing of the subpopulation dynamics in the patch property space, constituting a direct transformation that gives us the aggregated statistics from the patch-level variability.

\subsection{Deconvolution of the aggregated data}
\label{sec:deconvo}
In this section, I present the final step of the general approach, which involves the inverse transformation that can recover the patch variability from the aggregated statistics.
In order to avoid unnecessary complications, I will focus on a single patch property that belongs to a positive interval, $m\in \Omega \equiv [0,\infty)$, and positively influences population persistence. Such property could be resource availability, patch quality or dispersal rate, for instance.

Performing a change of variable in Eq.~(\ref{predirect}), setting $\tau = 1/s$ and writing $m(s)$, we obtain the integral in $s$
\begin{equation}
\mathcal{P}(T) = \mathcal{N}\int_{s(0)}^{s(\infty)} \frac{ \rho(m(s)) m'(s) s^2}{1 +  s \tau^{\dagger}(m(s))} e^{-T s} ds\, ,
\end{equation}
where prime notation means $m^\prime(s)=dm(s)/ds$. At this point, note that in order to write $m(s)$, the function set by the dynamics $s(m)$ needs to be invertible.

Since here it is assumed that the mean lifetime increases with the habitat property $m$, we can set the intervals to $s(\infty)=0$ and $s(0)=\infty$. Thus,
\begin{equation}
\label{superstat2}
\mathcal{P}(T) = -\mathcal{N}\int_0^{\infty} \frac{ \rho(m(s)) m'(s) s^2}{1 +  s \tau^{\dagger}(m(s))} e^{-T s} ds\, .
\end{equation}

Identifying that Eq.~(\ref{superstat2}) resembles the Laplace transform structure $\mathcal{L}[f](T) = \int_0^\infty f(s) e^{-T s} d s$, it is straightforward to obtain 
\begin{equation}
\label{direct}
\mathcal{P}(T) = -\mathcal{N}\mathcal{L} \left[\frac{ \rho(m(s)) m'(s) s^2}{1 +  s \tau^{\dagger}(m(s))} \right]\, .
\end{equation}

Finally, within the Laplace transform formalism, it is possible to apply the inverse transformation, recovering the distribution $\rho$ of property $m$ from the aggregated distribution $\mathcal{P}$,
\begin{align}
\rho(m) = - \mathcal{N}h(m)  \mathcal{L}^{-1}[\mathcal{P}(T)]_{s=1/\tau(m)} \, ,
\label{indirect}
\end{align}
where 
\begin{equation}
\label{hgeneral}
h(m) \equiv \frac{d \tau(m)}{dm}  \left(1 +  \frac{ \tau^{\dagger}(m)}{\tau(m)} \right)\, ,
\end{equation}
in which it was used that $s(m) = 1/\tau(m)$ and $ - s^2 m^\prime(s)= \frac{dm}{d\tau} $.

\section{Application}
\label{sec:apply}

Eqs.~(\ref{direct}) and (\ref{indirect}) provide us a two-way connection between patch-level variability ($\rho$) and the aggregated lifetime probability distribution ($P$). In order to show an application of this result, specific choices for how the mean extinction time $\tau$ depends on the patch property $m$ need to be made. 
I choose to set that $\tau$ grows exponentially with $m$,
\begin{equation}
\label{modelmigration}
\tau(m) = e^{a m} -1\, .
\end{equation}
For this choice, if $m$ is infinitely large, the population persists indefinitely. Conversely, when $m$ is negligible, the mean lifetime goes to zero. 

The proposed exponential form in Eq.~(\ref{modelmigration}) is motivated by the results found for when $m$ is the carrying capacity in standard population dynamics models, either by analytical or computational means~\cite{extinction}. This form is also observed when investigating the role of migration rates and patch connectivity, as reported in the supplementary material. For an effective single patch model, I show that its lifetime is exponentially distributed with mean lifetime that grows exponentially with the migration rate. For a spatial explicit formulation (under weak-coupling), considering a random patch landscape, analogous results were also observed, where the mean lifetime increases exponentially with connectivity. Thus, for the following results, the property $m$ can be interpreted as the patch carrying capacity, inward flux of individuals or connectivity, since all of them influence the mean lifetime in a similar way.

To specify the function $h$ in Eq.~(\ref{hgeneral}) it is also necessary to set how the empty time $\tau^\dagger$ depends on $m$. Particularly, I will set $\tau^\dagger(m) = 1/m$. This extra assumption regarding $\tau^\dagger$ applies for the scenarios in which $m$ refers to the migration rate (determining the rate of arrival of new individuals); or represents the carrying capacity when flux is biased towards high-quality patches (informed dispersal)~\cite{nos3,bengfort}. For the case in which $m$ do not couple to the migration process (uninformed dispersal), $\tau^\dagger$ will be constant, meaning homogeneous migration rates. For the following results, considering either $\tau^\dagger=1/m$ or constant does not produce qualitative changes.

Once the model mediation is set by electing $\tau(m)$ in Eq.~(\ref{modelmigration}), given the patch variability $\rho(m)$, it is possible to obtain the aggregated lifetime distribution $\mathcal{P}$ from Eq.~(\ref{direct}). More interestingly, it is possible to perform the reverse transformation (shown in Eq.~(\ref{indirect})), revealing the patch variability in the aggregated statistics.
So, I propose here a generalized power-law form for the aggregated lifetime probability distribution, namely
\begin{equation}
\label{stanleyform}
\mathcal{P}(T) \propto \frac{e^{-T/\tau_c}}{(\tau_0 + T)^{\beta}}\, ,
\end{equation}
where the parameters $\tau_c>0$ controls the exponential cut-off, and $\tau_0>0$ sets the probability of vanishing lifetimes and ensures normalization for the whole range $\beta>0$, assuming $T\in (0,\infty)$.
This choice is motivated by the North American Breeding Bird Survey dataset, for which was found $\beta=1.6$ and $\tau_c=14\text{ (years)}$~\cite{stanleybirds}, and were associated to critical phenomena~\cite{stanleybirds}. The following results aim to obtain the equivalent patch-level variability, that can generate same statistics at the aggregated level.

Substituting Eq.~(\ref{stanleyform}) in (\ref{indirect}), and noting that Eq.~(\ref{stanleyform}) has a common Laplace transform (i.e., the integrate unit step with frequency and time shifts), we finally obtain that
\begin{equation}
\rho(m) \propto h(m)[S(m)]^{\beta-1} e^{- S(m) \tau_0}\, ,\quad \text{for } m\in [0,m^\star]
\label{indirectapply}
\end{equation}
with $S(m) = 1/\tau(m) - 1/\tau_c$, being
\begin{equation}
\label{mstar}
m^\star = \frac{1}{a} \ln(\tau_c + 1)\, .
\end{equation} 

The function $h$ is obtained from Eq.~(\ref{hgeneral}), substituting Eq.~(\ref{modelmigration}), 
\begin{align}
h(m) = a e^{a m} \left(1 +  \frac{1}{m(e^{a m} -1)} \right) \, 
\label{happly}
\end{align}
where we additionally consider that $\tau^{\dagger}(m) = 1/m$, as discussed previously.

\begin{figure}[h]
\includegraphics[width=\columnwidth]{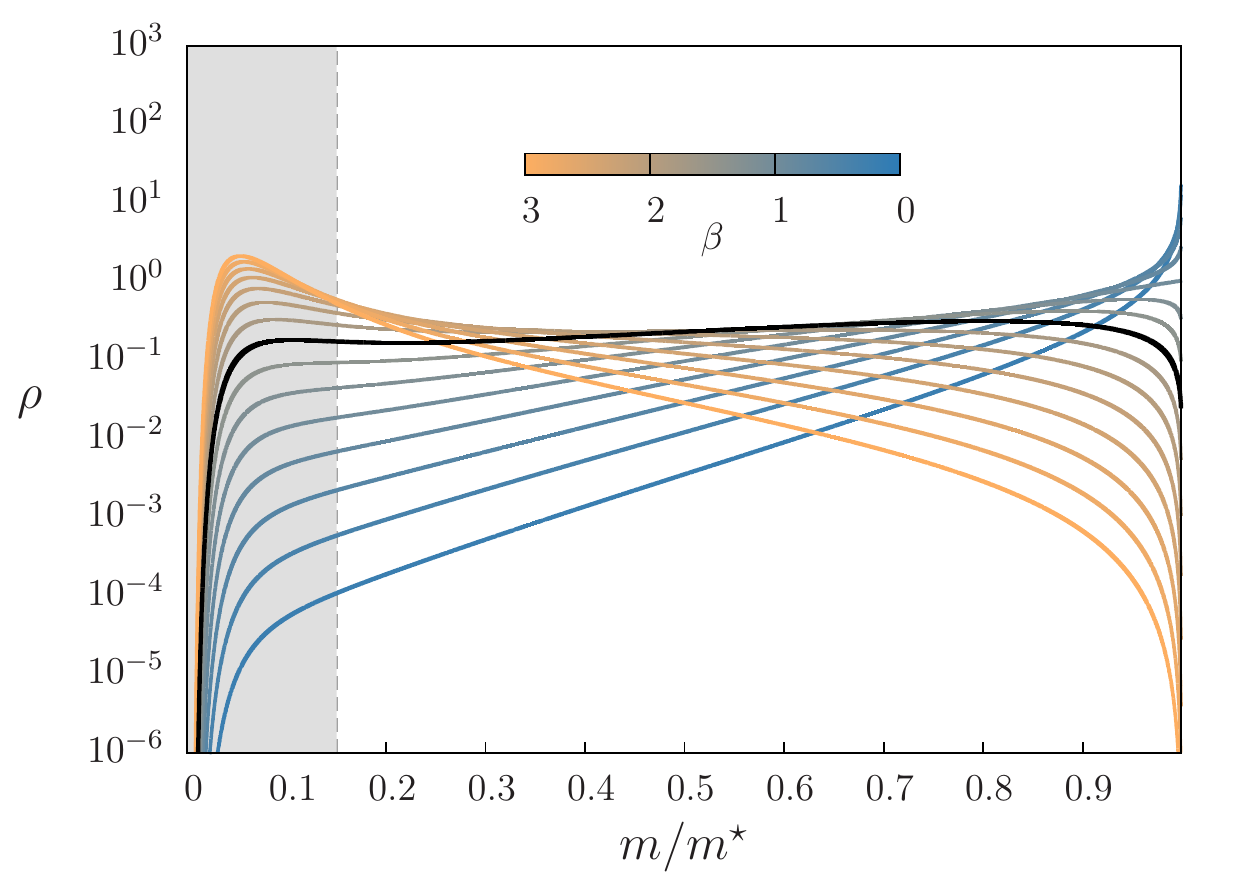}
\caption{Scaled probability distribution $\rho(m/m^\star)$ for different values of $\beta$. The distributions are obtained from Eq.~(\ref{indirectapply}) (with $a=1$, $\tau_0 = 1$, $\tau_c=100$) and $m^\star$ is given by Eq.~(\ref{mstar}). The gray region indicates the values of $m$ that correspond to mean lifetime $\tau<1$. The solid black line indicates the case $\beta=1.6$ from Ref.~\cite{stanleybirds}.}
\label{fig:stanley}
\end{figure}

In Fig.~\ref{fig:stanley}, it is shown how the probability distributions for patch variability $\rho(m)$ changes with the scaling exponent $\beta$ of Eq.~(\ref{stanleyform}).
Different values of the exponent reflects in qualitative differences in patch-level variability. For large $\beta$ values, the probability of observing patches with characteristic $m$ decays with the values of $m$, such that patches with high resource availability, migration rate or connectivity ($m \sim m^\star$) are rare in the system. In contrast, for small values of $\beta$ this behavior changes and the system exhibits a maximum at $m^\star$, being the wealthy patches (the ones with high connectivity and resource availability) numerous in the metapopulation. 

%%%

To identify the qualitative different distribution in Fig~\ref{fig:stanley}, it is helpful to analyze the behavior of $\rho$ in Eq.~(\ref{indirectapply}) under the limits of vanishing and maximum value of $m$.
For $m\to 0$, the mean lifetime becomes too small, being experimentally challenging to access this region. In Fig.~\ref{fig:stanley}, as a reference, the gray highlighted region corresponds to mean lifetime $\tau<1$, where the behavior of $\rho$ would not be captured by experimental data sampled in intervals with size $\Delta T \geq 1$ (every year, for example). Despite that, analytically, from Eq.~(\ref{indirectapply}), 
\begin{align}
\rho(0) \propto \frac{e^{-\frac{\tau_0}{am}}}{m^{\beta+1}} \biggr |_{m\to 0}\, .
\end{align}
Since, in Eq.~(\ref{stanleyform}) our proposal sets $\tau_0>0$ to ensure normalization, $\rho(m\to 0)$ always vanishes, as seen in Fig.~\ref{fig:stanley}. 

In the limit of maximum $m$ (high-quality limit), taking $ m \to m^\star$ in Eq.~(\ref{indirectapply}),
\begin{equation}
\label{lawstar}
\rho(m^\star) \propto (m^\star-m)^{\beta-1}\biggr |_{m\to m^\star}\, .
\end{equation}
Consequently, for $\beta>1$ ($\beta<1$), $\rho(m^\star)$ goes to zero (diverges), indicating the different regimes of metapopulation heterogeneity (as shown in Fig.~\ref{fig:stanley}). Only for $\beta=1$, $\rho(m^\star)$ assumes a positive finite value. Note that Eq.~(\ref{lawstar}) does not depend on other parameters aside from $\beta$.

\begin{figure}[ht]
\includegraphics[width=\columnwidth]{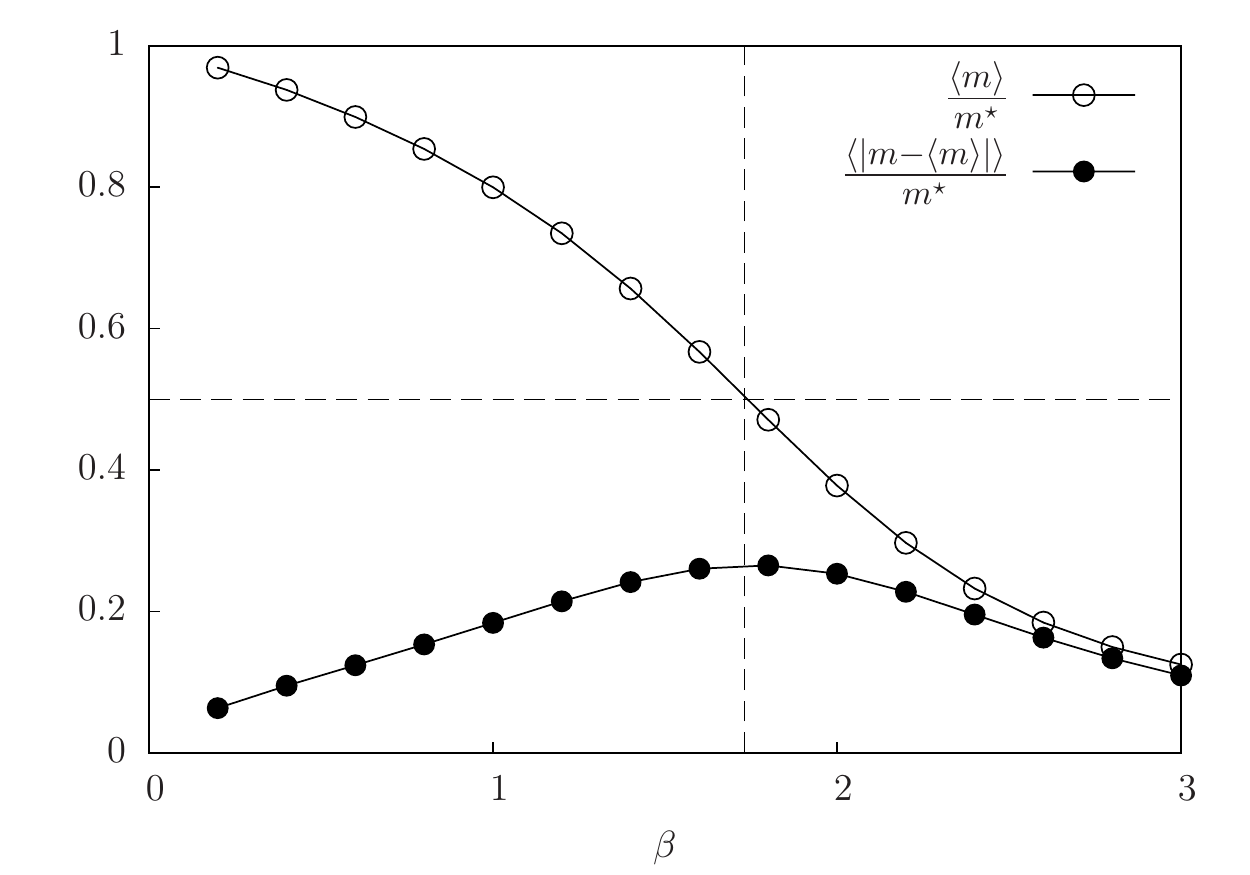}
\caption{Scaled mean and standard deviation of patch property $m$ as a function of $\beta$ for the distributions shown in Fig.~\ref{fig:stanley}. The horizontal dashed line indicates the line in which $\langle m \rangle/m^\star = 0.5$ and vertical dashed line shows the value of $\beta$ that produce maximum variability.}  
\label{fig:stats}
\end{figure}

More detailed information about the distributions shown in Fig.~\ref{fig:stanley} can be extracted obtaining the moments of the patch property distribution. In Fig.~\ref{fig:stats}, from the probability distribution given by Eq~(\ref{indirectapply}), I numerically compute the mean and the standard deviation of patch properties in the metapopulation for different values of $\beta$. As $\beta$ increases, the mean value of $m$ shifts towards small values, while the standard deviation passes through a maximum at $\beta \simeq 1.7$ (as indicated by the vertical dashed line). The particular value of the exponent that indicates maximum variability is sensitive to the parameters choice. Curiously, this optimal value of $\beta$ is close to the one observed in Ref.~\cite{stanleybirds}.

\section{Final remarks}

The origins of the power-laws observed in the aggregated lifetime probability distribution have been a matter of debate~\cite{stanleybirds,mixing}. On one hand, power-laws can be a consequence of critical phenomenon~\cite{stanleybook}, generating interconnected scaling laws for the system macroscopic state variable~\cite{stanleybook,andreas}. On the other hand, however, we see that patch-level variability can also control these power-laws.
During the previous sections, I studied the connection between the patch-level variability and the aggregated lifetime probability distribution. Considering a weakly-coupled metapopulation, the results demonstrate explicitly how patch variability can solely trigger different outcomes at the aggregated level. Understanding the intrinsic features of the dataset, I first wrote the aggregated lifetime probability distribution as a superstatistics~\cite{superstatistics} of the patches dynamics. Following that, an inverse transformation was obtained, revealing patch property variability from the aggregated perspective. 

Deepening previous debates~\cite{mixing,stanleybirds}, I applied these results to find the specific class of patch-level variability that generates scaling laws in the metapopulation aggregated lifetime probability distribution, mimicking the outcomes that can also be associated to critical phenomena~\cite{mixing}. Particularly, I investigated a generalized version of the power-law form observed for North American Breeding Bird Survey dataset~\cite{stanleybirds,databirds} which was the source of the debate in Ref.~\cite{mixing}. It was then presented that the observed scaling exponent $\beta$ is related to different types of heterogeneity. In the case of $\beta>1$, high-quality patches (well connected or with high resources availability) are rare in the metapopulation. While for $\beta<1$, this feature is inverted. Moreover, for a nontrivial value of $\beta$ ($\beta \simeq 1.7$, in the case depicted in Fig.~\ref{fig:stats}), patch variability achieves a maximum. The results have shown the connection between a macroscopic characteristic of the aggregated set ($\beta$) to patch-level variability ($\rho(m)$).

In sum, the application of the proposed framework indicates that, at the aggregated level, it is possible to have an equivalence between a heterogeneous weakly-coupled system and a homogeneous one exhibiting critical phenomena~\cite{mixing,lafuerza2013}. This stresses the role of patch-level variability in the aggregated data, which cannot be overlooked when aiming to interpret the underlying dynamics~\cite{mixing}.

%%%%
In order to achieve our results, some specific choices were made. Firstly, I assumed that the metapopulation is weakly coupled. On a first look, this approximation might be seen as a weakness of the results, however, it reveals its strength by demonstrating the emergence of power-laws in the aggregated lifetime statistics in the absence of strong correlations, solely triggered by patch heterogeneity~\cite{mixing,stanleybirds}. Secondly, the developments presented assumed that patch lifetimes are exponentially distributed. It was argued that this aspect is rather general, being present independently of the local population dynamics model. Nevertheless, it is worth to remark that in some special cases this feature might be violated at intermediate and short scales~\cite{fpt,extinction,effective}. Then, for these cases, the use of the Laplace transform formalism would be compromised.

It would be relevant to improve certain points of the results for practical situation. For instance, the calculations could be extended to the cases in which variability is present in several patch proprieties. However, to accomplish that, it is necessary to ensure the consistency of the steps done in the Sec.~\ref{sec:deconvo}. Also, fortunately, it was straightforward to apply the results to the aggregated probability distribution in the form of Eq.~(\ref{stanleyform}), since it has a common Laplace inverse. For special cases, one should develop a generalized transformation such as Mellin transformations~\cite{whittaker1996}.

At last, the results here developed can also be posed in a broader sense, as long as the dataset involves a collection of weakly correlated stochastic processes at different scales. For instance, in the context of animal movement, the aggregated probability distribution of the jump sizes is obtained by mixing data from random walkers that explore space at different spatial and temporal scales~\cite{foraging,tracking}. In this case, power-laws in movement statistics could emerge due to heterogeneity in animal behavior~\cite{stanleybirds,foraging,tracking}.

\section*{Supplementary material}

% Include only the SI item label in the paragraph heading. Use the \nameref{label} command to cite SI items in the text.
\paragraph*{S1 Appendix.}
\label{S1_Appendix}
{\bf Relation between lifetimes and patch properties.} Classical scenarios are investigated showing how mean population lifetime depends on patch properties.

\section*{Acknowledgments}
I would like to thank Profs. Emilio Hern\'andez-G\'arcia, Celia Anteneodo and Ra\'ul Toral for the support, comments and suggestions on the manuscript.
I also acknowledge funds from the Spanish Research Agency through grant ESOTECOS FIS2015-63628-C2-1-R (AEI/FEDER, EU) and grant MDM-2017-0711 from the Maria de Maeztu Program for units of Excellence in R\&D.

% Either type in your references using
% \begin{thebibliography}{}
% \bibitem{}
% Text
% \end{thebibliography}
%
% or
%
% Compile your BiBTeX database using our plos2015.bst
% style file and paste the contents of your .bbl file
% here. See http://journals.plos.org/plosone/s/latex for 
% step-by-step instructions.
% 
%\bibliographystyle{elsarticle-num}
%\bibliography{reference}

%

\end{document}